\documentclass[12pt]{article}
\pdfoutput=1 

\usepackage{graphicx}
\usepackage{amssymb}
\usepackage{epstopdf}
\usepackage{slashed}
\usepackage{float}
\usepackage{hyperref}
\usepackage{amsmath}
\usepackage{caption}
\usepackage{subcaption}
\usepackage{bbold}
\numberwithin{equation}{section}

\newcommand{\be}{\begin{equation}}
\newcommand{\ee}{\end{equation}}
\newcommand{\bea}{\begin{eqnarray}}
\newcommand{\eea}{\end{eqnarray}}
\newcommand{\beas}{\begin{eqnarray*}}
\newcommand{\eeas}{\end{eqnarray*}}
\newcommand{\ba}{\begin{array}}
\newcommand{\ea}{\end{array}}

\newcommand{\nn}{\nonumber}

\newcommand{\tr}[1]{\mbox{tr}\left( #1\right)}

\def\be{\begin{equation}}
\def\ee{\end{equation}}
\def\ben{\begin{equation*}}
\def\een{\end{equation*}}
\def\beqa{\begin{eqnarray}}
\def\eeqa{\end{eqnarray}}

\newcommand{\hA}{\hat{A}}

\newcommand{\m}{\mu}
\newcommand{\n}{\nu}

\newcommand{\g}{\gamma} 
\newcommand{\e}{\varepsilon}
\renewcommand{\d}{\partial}

\renewcommand{\(}{\left(}
\renewcommand{\)}{\right)}

\newcommand{\half}{\frac{1}{2}}

\newcommand{\dr}{\partial_r}

\newcommand{\dz}{\partial_z}

\newcommand{\dR}{\partial_R}
\newcommand{\dth}{\partial_\theta}

\newcommand{\Ah}{\hat{A}}

\def\bi{\begin{itemize}}
\def\ei{\end{itemize}}

\title{Holographic Baryons from Oblate Instantons}
\date{\today}
\author{Moshe Rozali, Jared B. Stang, Mark Van Raamsdonk\\ \\
\small Department of Physics and Astronomy, University of British Columbia\\
\small 6224 Agricultural Road, Vancouver, B.C., V6T 1Z1, Canada \\
\small \texttt{rozali, jstang, mav~\, @phas.ubc.ca} } 

\begin{document}
\maketitle

%====================================================================%
\begin{abstract}

We investigate properties of baryons in a family of holographic field theories related to the Sakai-Sugimoto model of holographic QCD. Starting with the $N_f=2$ Sakai-Sugimoto model, we truncate to a 5D Yang-Mills action for the gauge fields associated with the noncompact directions of the flavour D8-branes. We define a free parameter $\gamma$ that controls the strength of this Yang-Mills term relative to the Chern-Simons term that couples the Abelian gauge field to the $SU(2)$ instanton density. Moving away from $\gamma = 0$ should incorporate some of the effects of taking the Sakai-Sugimoto model away from large 't Hooft coupling $\lambda$. In this case, the baryon ground state corresponds to an oblate $SU(2)$ instanton on the bulk flavour branes: the usual $SO(4)$ symmetric instanton is deformed to spread more along the field theory directions than the radial direction. We numerically construct these anisotropic instanton solutions for various values of $\gamma$ and calculate the mass and baryon charge profile of the corresponding baryons. Using the value $\gamma = 2.55$ that has been found to best fit the mesonic spectrum of QCD, we find a value for the baryon mass of 1.19 GeV, significantly more realistic than the value 1.60 GeV computed previously using an $SO(4)$ symmetric ansatz for the instanton.

\end{abstract}
%====================================================================%

%====================================================================%
\newpage
\section{Introduction}

Perhaps the most successful holographic model of QCD has been the Sakai-Sugimoto model \cite{Sakai:2004cn,Sakai:2005yt}, defined by the physics of $N_f$ probe D8-branes in the background dual to the decoupling limit of $N_c$ D4-branes compactified on a circle with antiperiodic boundary conditions for the fermions. This model reproduces many features of real QCD, including chiral symmetry breaking, a deconfinement transition \cite{Witten:1998zw, Aharony:2006da}, and a realistic meson spectrum.

The description of baryons in the Sakai-Sugimoto model involves solitonic configurations of the Yang-Mills field on the D8-brane.\footnote{Mesons correspond to pertubative excitations of the D8-branes.} In a simplified ansatz where the Yang-Mills field is taken to depend only on the four non-compact spatial directions in the bulk, configurations with baryon charge are precisely those configurations with non-zero instanton number for this reduced 4D Yang-Mills field \cite{Sakai:2004cn, Hata:2007mb,Hong:2007kx,Hong:2007ay}. This connection between baryon charge and bulk instanton number stems from a Chern-Simons term $s \; \tr{F \wedge F}$ in the reduced D8-brane action. Here, $\tr{F \wedge F}$ is the instanton density for the $SU(2)$ part of the Yang-Mills field, and $s$ is the $U(1)$ part of the Yang-Mills field, dual to the baryon current operator in the field theory.

To date, the study of baryons in the Sakai-Sugimoto model has been somewhat unsatisfactory, for several reasons: I) While the action for the gauge field is of Born-Infeld type, only the leading Yang-Mills terms are typically used when studying the instantons. II)
For large 't Hooft coupling where the model can be studied most reliably, the size of the instanton in the bulk has been argued to be much smaller than the size of the compact directions in the bulk. In this case, the assumption that the gauge field does not depend on the compact directions is questionable. III) Rather than solving the bulk equations to determine the precise solitonic configuration of the Yang-Mills field, the form has been taken to be that of a flat-space $SO(4)$ symmetric instanton, with the size of the instanton as the only free parameter.

The assumptions in I) and II) here amount to replacing the original top-down Sakai-Sugimoto model with a phenomenological (bottom-up) holographic model that retains many of the same successes as the Sakai-Sugimoto model. For the present paper, we continue to make these assumptions, though we hope to relax them in future work in order to better understand baryons in the fully-consistent top-down model. Our goal in the present paper is to overcome the third deficiency, by setting up and solving numerically a set of partial differential equations that determine the proper form of the soliton.\footnote{\cite{Pomarol:2007kr,Pomarol:2008aa,Pomarol:2009hp} have used a similar numerical approach in other phenomenological holographic QCD models.}  Using these solutions, we are able to calculate the mass and baryon charge distribution of the baryons as a function of the model parameter $\gamma$ (proportional to the inverse 't Hooft coupling $\lambda$) that controls the strength of the Chern-Simons term relative to the Yang-Mills term.

One motivation for our work is the work of \cite{Cherman:2009gb}, which points out that the flat-space instanton approximation used previously does not give the correct large radius asymptotic behavior (known from model-independent constraints) for the baryon form factors (computed for example in \cite{Hong:2007dq,Hashimoto:2008zw,Kim:2008pw}). Via a perturbative expansion of the equations at large radius, it was later shown \cite{Cherman:2011ve} that by relaxing the assumption of $SO(4)$ symmetry, the proper asymptotic behavior can be recovered.\footnote{In the earlier work \cite{Panico:2008it}, a similar expansion was used in a phenomenological holographic QCD model. See also \cite{Colangelo:2013pxk} for a recent related study.} Thus, we expect that by constructing and studying the complete solutions, we can obtain a significantly improved picture of the properties of baryons in holographic QCD.

The solutions that we find take the form of ``oblate instantons'': compared with the $SO(4)$ symmetric configurations, the correct solutions are deformed to configurations with $SO(3)$ symmetry that are spread out more in the field theory directions than in the radial direction. This shape is expected. The Coulomb repulsion between instanton charge density at different locations (induced by the Chern-Simons coupling to the Abelian gauge field) acts symmetrically in all directions, impelling the instanton to spread out both in the radial and field theory directions. Gravitational forces in the bulk limit the spreading in the radial direction, but there are no equivalent forces acting to radially compress the instanton in the field theory directions. Thus, the instanton is oblate, compressed in one direction relative to the other three. The anisotropy is limited by the Yang-Mills action for the $SU(2)$ gauge field, which in flat space is minimized (in the one-instanton sector) for spherically symmetric configurations.

The size and anisotropy of the instantons is controlled by the parameter  $\gamma$ (related to the inverse 't Hooft coupling in the original model). For small $\gamma$, the spreading effects of the Chern-Simons term are small, and the instantons become small and approximately symmetrical near their core. For larger $\gamma$, the instantons become significantly larger and more anisotropic. Using our numerics, we are able to construct solutions up to $\gamma$ of order 100 and evaluate the mass and baryon charge profiles of the corresponding baryons.

While our model is not expected to quantitatively match real-world QCD measurements, previous studies have found that the meson spectrum agrees reasonably well with the spectrum in QCD for a suitable choice of the parameter $\gamma$. Thus, it is interesting to compare the mass and size of the baryons in our model to the QCD values for the light nucleons. Using the value $\gamma = 2.55$ that gives the best fit to the meson spectrum \cite{Hashimoto:2008zw}, we find that the mass and baryon charge radius of the baryon are 1.19 GeV and 0.90 fm. This mass is significantly closer to realistic values ($\sim$ 0.94 GeV for the proton and neutron) than the previous value of 1.60 GeV based on the $SO(4)$ symmetric ansatz. The baryon charge radius is quite similar to measured values for the size of the proton and neutron. For example, the electric charge radius of a proton has been measured to be in the range 0.84 fm -- 0.88 fm \cite{Beringer:1900zz}, while the magnetic radii of the proton and neutron are listed in \cite{Beringer:1900zz} as 0.78 fm and 0.86 fm respectively.

An outline for the remainder of the paper is as follows:
In section \ref{sec:SS}, we briefly review the description of baryons in the Sakai-Sugimoto model and set up the problem. In section \ref{sec:Num}, we describe our numerical approach to the equations. In section \ref{sec:Solutions}, we describe physical properties of the solution, focusing on the baryon mass and the distribution of baryon charge (charge density as a function of radius), as a function of $\gamma$. Our main results may be found in figures \ref{Mass} and \ref{RhoB1}. We conclude in section {\ref{sec:Conc}} with a brief discussion of directions for future work.

Note: While this work was being completed, \cite{Bolognesi:2013nja} appeared, which also presents a numerical solution of the Sakai-Sugimoto $N_B=1$ soliton, using different methods, and which has some overlap with this paper.

%====================================================================%
\section{Baryons as solitons in the Sakai-Sugimoto model}
\label{sec:SS}

In this section, we give a brief review of the Sakai-Sugimoto model and set up the construction of a baryon in this model.

The Sakai-Sugimoto model consists of $N_f$ probe D8 branes in the near horizon geometry of $N_c$ D4 branes wrapped on a circle with anti-periodic boundary conditions for the fermions. The metric of the D4 background is \cite{Witten:1998zw}
\begin{align}ds^2& = {\lambda \over 3}  l_s^2 \(\frac{4}{9} u^{3 \over 2} \(\eta_{\m \n} dx^\m dx^\n +f(u)d x_4^2\) + \frac{1}{u^{3 \over 2}} \(\frac{du^2}{f(u)}+u^2 d\Omega^2_4\)\), \nn\\
e^\Phi&=\(\frac{\lambda}{3}\)^{3 \over 2}\frac{u^{3 \over 4}}{\pi N_c},\quad f(u)=1-\frac{1}{u^3},\quad F_4=dC_3=\frac{2\pi N_c}{V_4}\epsilon_4, \label{eq:bg}
\end{align}
where $\epsilon_4$ is the volume form on $S^4$ and $V_4$ is the volume of the unit 4-sphere. The direction $x_4$, with radius $2\pi$, corresponds to the direction on which the D4-branes are compactified. The $u$ and $x_4$ directions form a cigar-type geometry and the space pinches off at $u=1$. The four dimensional $SU(N_c)$ gauge theory dual to this metric has a dimensionless coupling $\lambda$.

The flavor degrees of freedom are provided by $N_f$ probe D8 branes in the background (\ref{eq:bg}). The action for a single D8 brane is \be S_{D8}=-\m_8\int d^9\sigma e^{-\Phi}\sqrt{-\det(g_{ab}+2\pi\alpha'\mathcal{F}_{ab})}+S_{CS},\label{eq:DBI}\ee with $\m_8=1/(2\pi)^8l^9_s$ and where $S_{CS}$ is the Chern-Simons term. Below, we expand this action around a particular embedding and take the non-Abelian generalization of the result to define the action we consider. We take the probe branes to wrap the sphere directions and fill the $3+1$ field theory directions. Then, the embedding is described by a curve $x_4(u)$ in the cigar geometry, with boundary conditions fixing the position of the probe branes as $u \rightarrow \infty$.

 In this paper, we consider only the antipodal case, in which the ends of the probe branes are held at opposite sides of the $x_4$ circle. The minimum energy configuration with these boundary conditions is that in which the probe branes extend down the cigar at constant angle $x_4$, meeting at $u=1$. Going to the radial coordinate $z$ defined by $u^3=1 + z^2$, and expanding the action (\ref{eq:DBI}) for small gauge fields around the antipodal embedding gives the model we consider \cite{Hata:2007mb}:
\begin{align} S=-\kappa\int d^4xdz \;\textrm{tr}&\left[\half h(z)\mathcal{F}^2_{\mu\nu} + k(z)\mathcal{F}^2_{\mu z} \right]\nn\\
&+\frac{N_c}{24\pi^2}\int_{M_5} \mathrm{tr} \left( \mathcal{A} \, \mathcal{F}^2 - \frac{i}{2} \mathcal{A}^3 \mathcal{F} - \frac{1}{10}\mathcal{A}^5 \right),\label{eq:S}
\end{align} where $\kappa=\lambda N_c/(216\pi^3)$, $h(z)=(1+z^2)^{-1/3}$ and $k(z)=1+z^2$. $\mathcal{A}$ is a $U(N_f)$ gauge field with field strength $\mathcal{F}=d\mathcal{A}+i\mathcal{A}\wedge\mathcal{A}$. In this paper, we focus on the case $N_f=2$. We split the gauge field into $SU(2)$ and $U(1)$ parts as $\mathcal{A}=A+\half \mathbb{1}_2 \Ah$.\footnote{We define the $SU(2)$ generators to satisfy $[\tau^a,\tau^b]=i\e^{abc}\tau^c$.}

The competing forces that determine the size of the soliton are evident in the effective action (\ref{eq:S}). First, the gravitational potential of the curved background will work to localize the soliton near the tip of the cigar, at $z=0$. This will be counterbalanced by the repulsive potential due to the coupling between the $U(1)$ part of the gauge field and the instanton charge in the Chern-Simons term. At large $\lambda$, the effect of the Chern-Simons term is suppressed, and the result is a small instanton, which was previously approximated by the flat-space $SO(4)$ symmetric BPST instanton. As discussed in \cite{Cherman:2011ve}, this approach fails to properly describe several aspects of the baryon. Due to the curved background, the actual solution will only be invariant under $SO(3)$ rotations in the field theory directions. This distinction is especially important if we wish to use this model away from the strict large $\lambda$ limit, as in that case, the soliton can become large such that the effects of the curved background are important for more than just the asymptotics of the solution.

The most general field configuration invariant under combined $SO(3)$ rotations and $SU(2)$ gauge transformations may be written as \cite{Witten:1976ck,Forgacs}\footnote{This ansatz has also been used in the study of holographic QCD in a phenomenological model \cite{Pomarol:2007kr,Pomarol:2008aa,Pomarol:2009hp} and was applied to the Sakai-Sugimoto model in \cite{Cherman:2011ve}.}
\begin{align}
\label{eq:Ansatz}
A_{j}^{a} &= \frac{\phi_{2}+1}{r^{2}} \epsilon_{j a k} x_{k} + \frac{\phi_{1}}{r^{3}} [\delta_{ja} r^{2} - x_{j} x_{a}]+A_{r} \frac{x_{j}x_{a}}{r^{2}}, \nn \\
A_{z}^{a} &= A_{z} \frac{x^{a}}{r},  \quad \hA_{0} = \hat{s}.
\end{align}
where each of the fields are functions of the boundary radial coordinate $r=x^ax^a$ and the holographic radial coordinate $z$. The ranges of these coordinates are $0<r<\infty$ and $-\infty<z<\infty$. With these definitions, there is a residual gauge symmetry under which $A_\m$ transforms as a $U(1)$ gauge field in the $r-z$ plane and $\phi=\phi_1+i\phi_2$ transforms as a complex scalar field with charge $(-1)$, so that $D_\m\phi=\d_\mu\phi-iA_\mu\phi$.

The free energy of the system is given by the Euclidean action evaluated on the solution. Since we work at zero temperature and consider only static solutions, the mass-energy equals the free energy, and we only pick up a minus sign from the analytic continuation. Then, in terms of the above ansatz, the mass of the system is written as
\begin{align}
\label{eq:Mass}
M=M_{YM}+M_{CS} ,
\end{align}
where $\int dt M = -S$,
\begin{align}
M_{YM} = 4 \pi \kappa \int dr dz \,  &\left[  h(z) |D_{r} \phi|^{2} + k(z) |D_{z} \phi|^{2}  +\frac{1}{4}r^{2} k(z) F_{\mu\nu}^{2} \right.\nn\\
&\left. +\frac{1}{2r^{2}} h(z) (1-|\phi|^{2})^{2} -\frac{1}{2}r^{2} \left(h(z) (\partial_{r} \hat{s})^{2} + k(z) (\partial_{z} \hat{s})^{2}\right) \right]
\end{align}
and
\begin{align}
M_{CS} =-2\pi \kappa \gamma \int dr dz \, \hat{s}\,\epsilon^{\mu \nu}\left[ \partial_{\mu}(-i\phi^{*}D_{\nu}\phi + h.c.) +F_{\mu \nu}\right] ,
\end{align} 
with $\gamma = N_{c}/(16\pi^{2} \kappa) = 27\pi/(2\lambda)$ and $F_{\m\n}=\d_\m A_\n-\d_\n A_\m$. For the classical solution, $\g$ is the only parameter in the system. It controls the relative strength of the Chern-Simons term; a larger $\g$ will increase the size of the soliton.

The equations of motion that follow from extremizing the mass-energy are given by
\begin{align}
0 &= D_{r} \left(h(z) D_{r} \phi \right)+D_{z} \left(k(z) D_{z} \phi \right) + \frac{h(z)}{r^{2}} \phi(1-|\phi|^{2}) +i \gamma \epsilon^{\mu \nu} \partial_{\mu} \hat{s}  D_{\nu} \phi , \nn\\
0 &=\partial_{r} \left(r^{2} k(z) F_{r z}\right) - k(z) \left(i \phi^{\ast} D_{z}\phi + h.c. \right) - \gamma \epsilon^{r z} \partial_{r}\hat{s}(1-|\phi|^{2}) , \nn\\
0 &=\partial_{z} \left(r^{2} k(z) F_{z r}\right) - h(z) \left(i \phi^{\ast} D_{r}\phi + h.c. \right) - \gamma \epsilon^{z r} \partial_{z}\hat{s}(1-|\phi|^{2}) , \nn\\
0 &=\partial_{r} \left( h(z) r^{2} \partial_{r} \hat{s} \right) +\partial_{z} \left( k(z) r^{2} \partial_{z} \hat{s} \right)  - \frac{\gamma}{2} \epsilon^{\mu \nu} \left[ \partial_{\mu}(-i\phi^{\ast}D_{\nu}\phi + h.c) +F_{\mu \nu}\right].
\end{align}

The baryon number is given by the instanton number of the non-Abelian part of the gauge field,
\bea N_B &=& \frac{1}{8\pi^2} \int d^4x\, \mathrm{tr}\,F\wedge F\nn\\
&=&\frac{1}{4\pi} \int drdz\, \epsilon^{\mu \nu}\left[ \partial_{\mu}(-i\phi^{*}D_{\nu}\phi + h.c.) +F_{\mu \nu}\right] \nn\\
&=&\frac{1}{4\pi} \int drdz\, (\dr q_r + \dz q_z), \label{eq:NB1}
\eea
where $F$ is the field strength of the $SU(2)$ gauge field $A$ and
\be q_r = (-i\phi^{*}D_{z}\phi + h.c.) + 2A_z,\quad q_z = (i\phi^{*}D_{r}\phi + h.c.) - 2A_r. \ee
Since the expression is a total derivative, the boundary conditions on our $SU(2)$ gauge field will set the baryon charge. We study configurations with $N_B=1$.

%====================================================================%
\section{Numerical setup and boundary conditions}
\label{sec:Num}

In this section we describe our setup, including our boundary conditions, gauge fixing and details about the numerical procedure we use.

\subsection{Gauge fixing}
\label{gauge}

There is a residual $U(1)$ gauge freedom in the above ansatz, and we choose to use the Lorentz gauge $\chi \equiv \d_\m A_\m=0$. Our gauge fixing is achieved by adding a gauge fixing term to the equations of motion, analogous to the Einstein-DeTurck method developed in \cite{Headrick:2009pv}. Alternatively, one can view this procedure as adding a gauge fixing term to the action, and working in the Feynman gauge.

As a result one obtains modified equations of motion in which the principal part of the equations is simply the standard elliptic operator $\partial_r^2+\partial_z^2$. Once a solution is obtained, one has to make sure it is also a solution to the original, unmodified  equations, i.e that $\chi=0$. This has to be checked numerically, but can be expected to be satisfied since $\chi$ is a harmonic function, so with suitably chosen boundary conditions (for example such that $\chi=0$ on the boundaries of the integration domain) uniqueness of solution to Laplace equation guarantees that $\chi=0$. For the solutions presented here, the gauge condition is well satisfied as the $L^2$ norm of $\chi$, normalized by the number of grid points $N$, satisfies $|\chi|/N<10^{-5}$.

\subsection{Ansatz and boundary conditions}

For small $\g$, the soliton solution is well localized near the origin $(r,z)=(0,0)$. For small $z$, $k(z)\sim h(z)\sim1$ and the $SU(2)$ part of the action reduces to that of the Witten model \cite{Witten:1976ck} for instantons. Then, in this regime, we expect the solution to possess an approximate $SO(4)$ symmetry, and thus we find it convenient to use the spherical coordinates \be R=\sqrt{r^2+z^2}, \quad \theta=\arctan(r/z) \ee for our numerical calculation. The inverse transformation is $r=R\sin\theta,\quad z=R\cos\theta$. One can show that by restricting the ansatz (\ref{eq:Ansatz}) to $SO(4)$ symmetry,\footnote{This assumption would be valid if $k$ and $h$ were spherically symmetric. The Chern-Simons term does not break the $SO(4)$ symmetry.} the solution can be written in terms of two spherically symmetric functions $f(R)$ and $g(R)$ as
\be \label{eq:SO4Ansatz} \phi_1=-rzf(R),\quad \phi_2=r^2f(R)-1, \quad A_r=-zf(R),\quad A_z=rf(R), \quad \hat{s}=g(R). \ee
In this parametrization, the BPST instanton is given by \be f(R) = \frac{2}{\rho^2+R^2},\quad g(R)=0, \ee where $\rho$ determines the size of the energy distribution. The non-trivial winding of the instanton is built into the expressions in (\ref{eq:SO4Ansatz}) through the appropriate factors of $r$ and $z$ and the factor of $2$ in the numerator of $f(R)$ fixes the winding number to be $N_B=1$. The BPST solution has a scaling symmetry in that it admits solutions of arbitrary scale $\rho$.

The factors of $k(z)$ and $h(z)$ in the Sakai-Sugimoto model break the $SO(4)$ symmetry. This has two effects on the $SO(4)$ ansatz. First, the functions $\phi_1,\phi_2,A_r$ and $A_z$ will not be related to each other through the common function $f(R)$. Second, the functions appearing in the ansatz must be promoted to functions of both the radial coordinate $R$ and the angle $\theta$. These considerations motivate our reduced ansatz as
\begin{align} \label{eq:NumAnsatz}
\phi_1&=-\(\frac{R^2\sin\theta\cos\theta}{1+R^2}\)\psi_1(R,\theta),\quad \phi_2=\(\frac{R^2\sin^2\theta}{1+R^2}\)\psi_2(R,\theta)-1,\nn\\
A_r&=-\(\frac{R\cos\theta}{1+R^2}\)a_r(R,\theta),\quad A_z=\(\frac{R\sin\theta}{1+R^2}\)a_z(R,\theta), \quad \hat{s}=\frac{s(R,\theta)}{R\sin\theta}.
\end{align}
In each of the non-Abelian gauge field functions we include a factor of $(1+R^2)^{-1}$ such that we may use Dirichlet boundary conditions at $R=\infty$ to fix the baryon number. We rescale $s$ by a factor of $r^{-1}=(R\sin\theta)^{-1}$ in order to have better control over the behaviour of the gauge field near the $r=0$ boundary. We numerically solve for the five functions $\{\psi_1,\psi_2,a_r,a_z,s\}$ on the domain $(0\leq R<\infty, 0\leq\theta\leq\pi/2)$ corresponding to $(0\leq r<\infty,0\leq z<\infty)$. In practice, we use a finite cutoff at $R=R_\infty$, chosen such that the physical data extracted from the solution does not depend on it. The symmetries of the solution around $z=0$ are used to extend it to $(-\infty<z\leq0)$.

In terms of the coordinates $(R,\theta)$, the baryon charge becomes \be N_B = \frac{1}{4\pi} \int dRd\theta\, (\dR q_R + \dth q_\theta), \ee
where we have defined \be q_R=R(\sin\theta q_r + \cos\theta q_z), \quad q_\theta= \cos\theta q_r-\sin\theta q_z.\ee The baryon number is given by the boundary integrals \bea N_B=\frac{1}{4\pi}\(\int_0^\infty dR\, q_\theta\Big|_{\theta=0} + \int_0^\pi d\theta\, q_R\Big|_{R=\infty} + \int_\infty^0 dR\, q_\theta\Big|_{\theta=\pi} + \int_\pi^0 d\theta\, q_R\Big|_{R=0}\). \eea Plugging our ansatz into $q_R$ and $q_\theta$ and evaluating on the boundaries shows that the only contribution to the winding is from the boundary at $R=\infty$. Thus, the baryon number reduces to \be N_B=\frac{1}{2\pi} \int_0^{\pi/2} d\theta\, q_R\Big|_{R=\infty}, \ee and we use boundary conditions at the cutoff $R_\infty$ to impose that $N_B=1$.

The boundary conditions we use are as follows. At $\theta=\pi/2$ (which maps back to $z=0$), we have Neumann conditions on all the fields, as the odd/even characteristics of the functions about $z=0$ are built into the ansatz (\ref{eq:NumAnsatz}). At this boundary $\chi=0$ implies $\dth a_z=0$ so that this boundary condition satisfies the gauge choice. To obtain boundary conditions at $\theta=0$ ($r=0$), we expand the equations of motion for small $\theta$. Satisfying these order by order in $\theta$ gives a set of conditions on the fields. A subset of these conditions that results in a convergent solution is given by\footnote{In practice, we use the boundary condition $\dth a_z=\half R\,\dth\dR\psi_1$ during the solving procedure, as we found empirically that this results in a more stable Newton iteration. Once the numerical procedure converges, the solution satisfies the boundary conditions given here.} \be \theta=0:\quad\dth\psi_1=0,\quad \dth\psi_2=0,\quad a_r=\psi_1,\quad \dth a_z=0,\quad s=0.\quad\ee The gauge condition at $\theta=0$ can be shown to be satisfied on a solution given these boundary conditions. At the origin $R=0$, a similar procedure yields \be R=0:\quad \dR\psi_1=0,\quad \dR\psi_2=0,\quad \dR a_r=0,\quad \dR a_z=0,\quad s=0. \ee We do not explicitly satisfy the gauge condition $R=0$.\footnote{We check that the gauge condition $\chi=0$ is numerically satisfied on our solutions across the domain. See section \ref{gauge}.} At the cutoff $R_\infty$, the boundary conditions are determined by behaviour of the gauge field $\hA_0$ and the winding number $N_B=1$. As discussed below, in section \ref{sec:RhoB}, the field theory density of baryon charge $\rho_B(r)$ (defined below) is proportional to the coefficient of the $z^{-1}$ falloff of the Abelian gauge field $\hA_0$, at large $z$. In order to reliably calculate $\rho_B(r)$, we therefore impose that $s$ falls off as $z^{-1}$ by using the boundary condition $s=-z\dz s$, suitably translated into $(R,\theta)$ coordinates, at the cutoff $R_\infty$. Since we rescaled the $SU(2)$ gauge fields by $(1+R^2)^{-1}$, we are left with Dirichlet conditions on the other functions, giving \be R=R_\infty:\quad \psi_1=\psi_2=a_r=a_z=2, \quad s=-R\cos^2\theta\,\dR s+\sin\theta\cos\theta\,\dth s.\ee Given the asymptotic boundary behavior of the fields, the gauge choice is satisfied for large $R_\infty$. With these large $R$ conditions, we have $q_R=4$ and so $N_B=1$, as desired.

\subsection{Numerical procedure}
\begin{figure}[htb!]
\centering%
\includegraphics[scale=1.5]{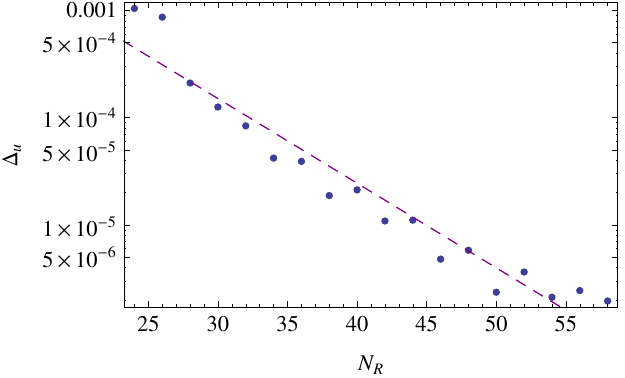}
\caption{The convergence of the value $\Delta_u=|u(N_R)-u(N_R-2)|/N_RN_\theta$, where $u(N_R)$ denotes the solution for the five fields $\{\psi_1,\psi_2,a_r,a_z,s\}$ on the grid with $N_R$ points in the $R$ direction and $N_\theta=N_R/2$ points in the $\theta$ direction. These runs are for $\g=10$ and $R_\infty=60$. The dashed line is the best linear fit, showing the exponential convergence $\Delta_u\propto e^{-0.18N}$.}
\label{convergence}
\end{figure}

We solve the equations of motion by using spectral methods on a Chebyshev grid, using the Newton method to solve the resulting non-linear algebraic equations. For the results presented here, we take the number of grid points to be $(N_R,N_\theta)=(50,25)$. We introduce a cutoff at large $R=R_{\infty}$. For a large enough cutoff we can reliably read off the $z^{-1}$ falloff in order to obtain information about the baryon charge density. However, if the cutoff is too large, the total mass-energy of the solution becomes dependent on $R_\infty$. In practice, we take $R_\infty$ to vary with $\g$, such that we can compute both the mass-energy and the baryon charge density with confidence across most of our domain. We find that while the charge density can be computed to good accuracy for large $\g$, the mass-energy becomes unreliable for $\g\gtrsim70$. To generate a solution, we continue the Newton method until the residuals reach a very small value ($\sim10^{-9}$). For generic values of $\g$, we can solve for the configuration from a trivial initial guess (zero for all the fields), while for very large or very small $\g$, we solve by using a nearby solution as the initial guess. Finally, the convergence of our solutions is demonstrated in Figure \ref{convergence}.

%====================================================================%
\section{Solutions}
\label{sec:Solutions}

We focus on two observables of the baryon in the Sakai-Sugimoto model: the mass-energy and the baryon charge density. We examine each of these in turn.

\subsection{The mass-energy}

\begin{figure}
\centering
\begin{subfigure}{.45\textwidth}
  \centering
  \includegraphics[width=1\linewidth]{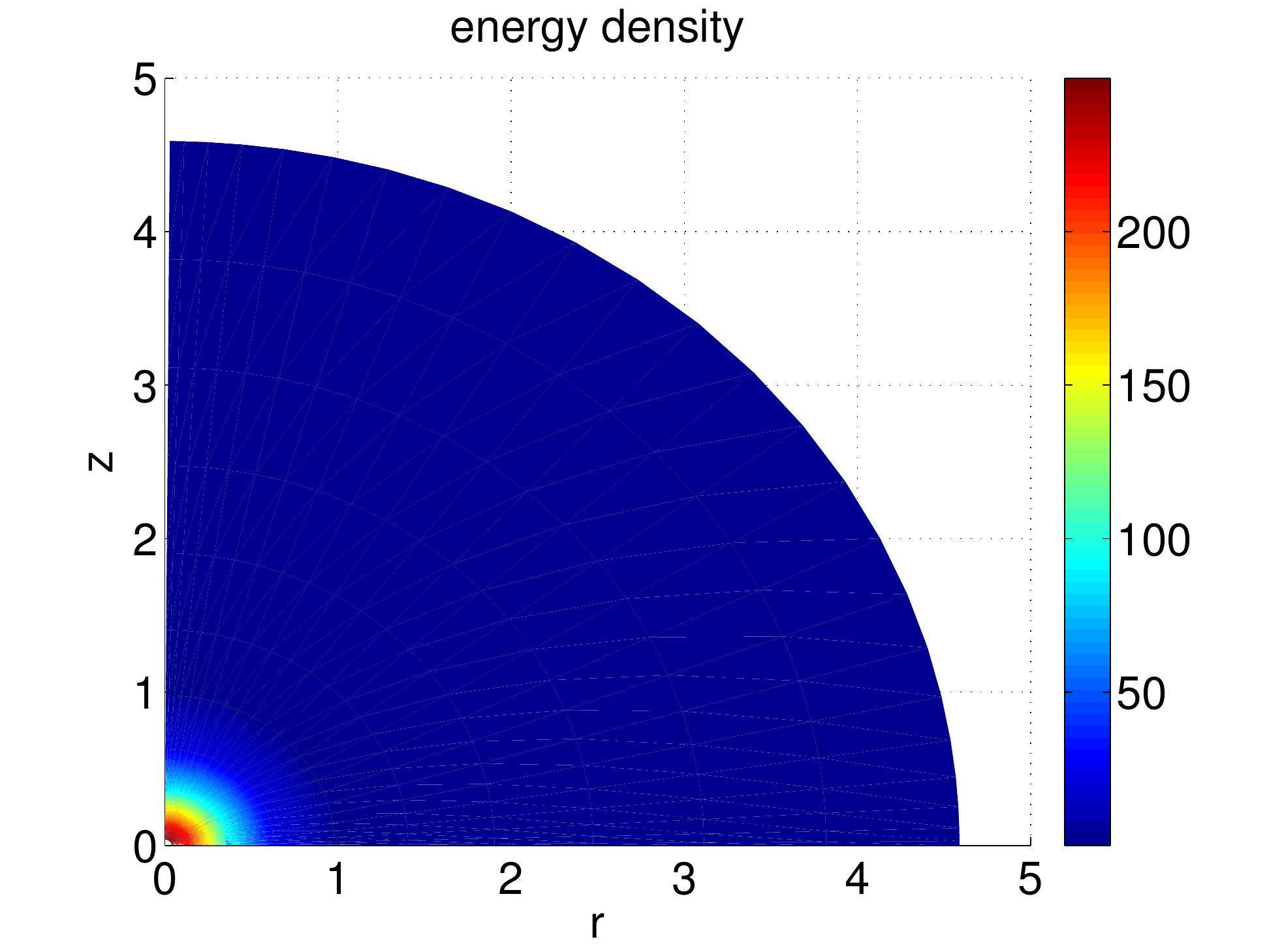}
  \caption{$\g=0.2$.} % Actual gamma (as defined in paper) is twice the saved value.
  \label{RhoE_02}
\end{subfigure}
\begin{subfigure}{.45\textwidth}
  \centering
  \includegraphics[width=1\linewidth]{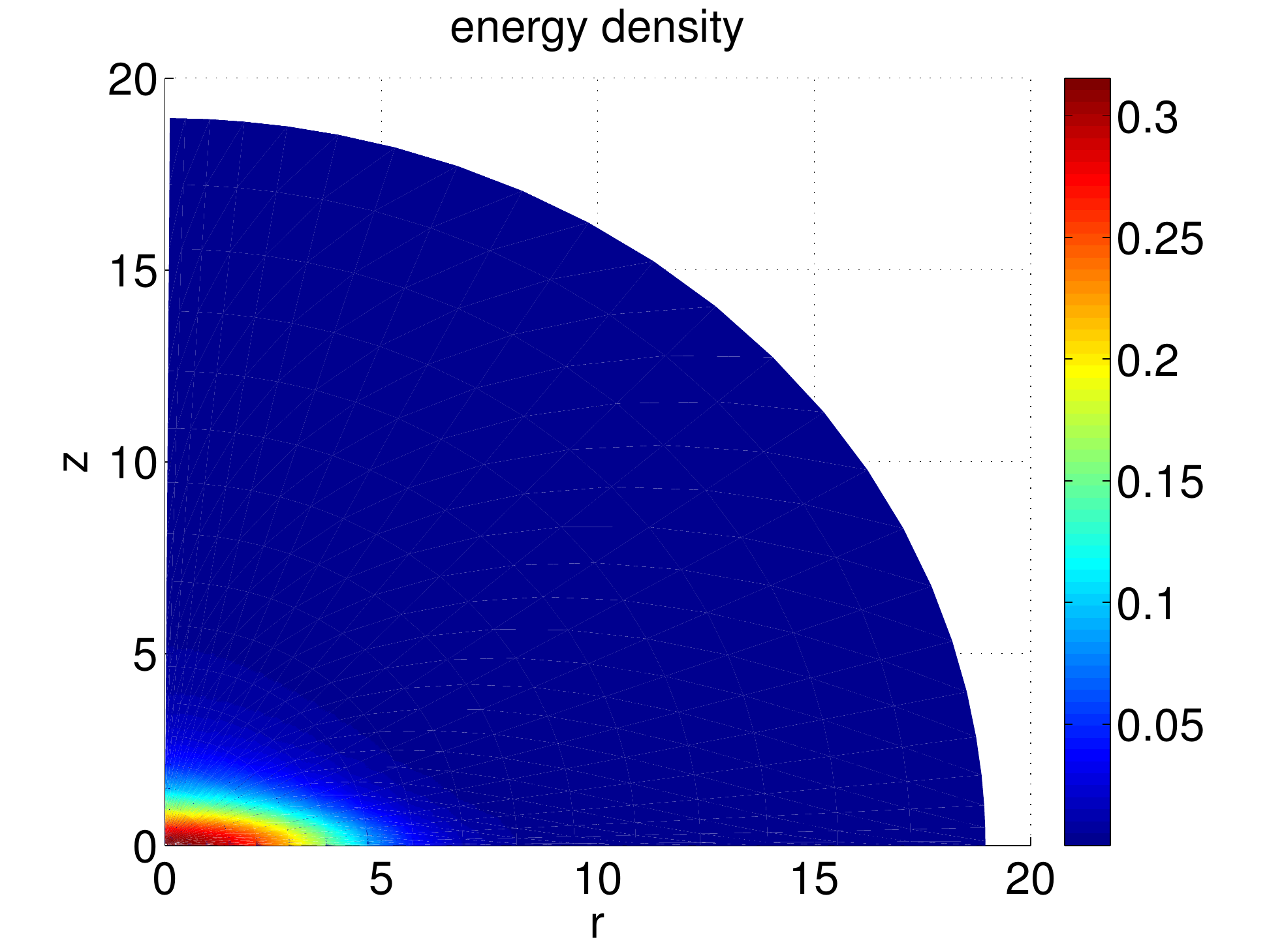}
  \caption{$\g=10$.}
  \label{RhoE_10}
\end{subfigure}
\caption{The energy density $\rho_E(r,z)$ in the $(r,z)$ plane. For small $\g$, the solution appears approximately spherically symmetric. As the coupling $\g$ increases, the soliton expands and deforms, becoming elongated along $z=0$.}
\label{RhoE_Gamma}
\end{figure}

\begin{figure}
\centering
\begin{subfigure}{.45\textwidth}
  \centering
  \includegraphics[width=1\linewidth]{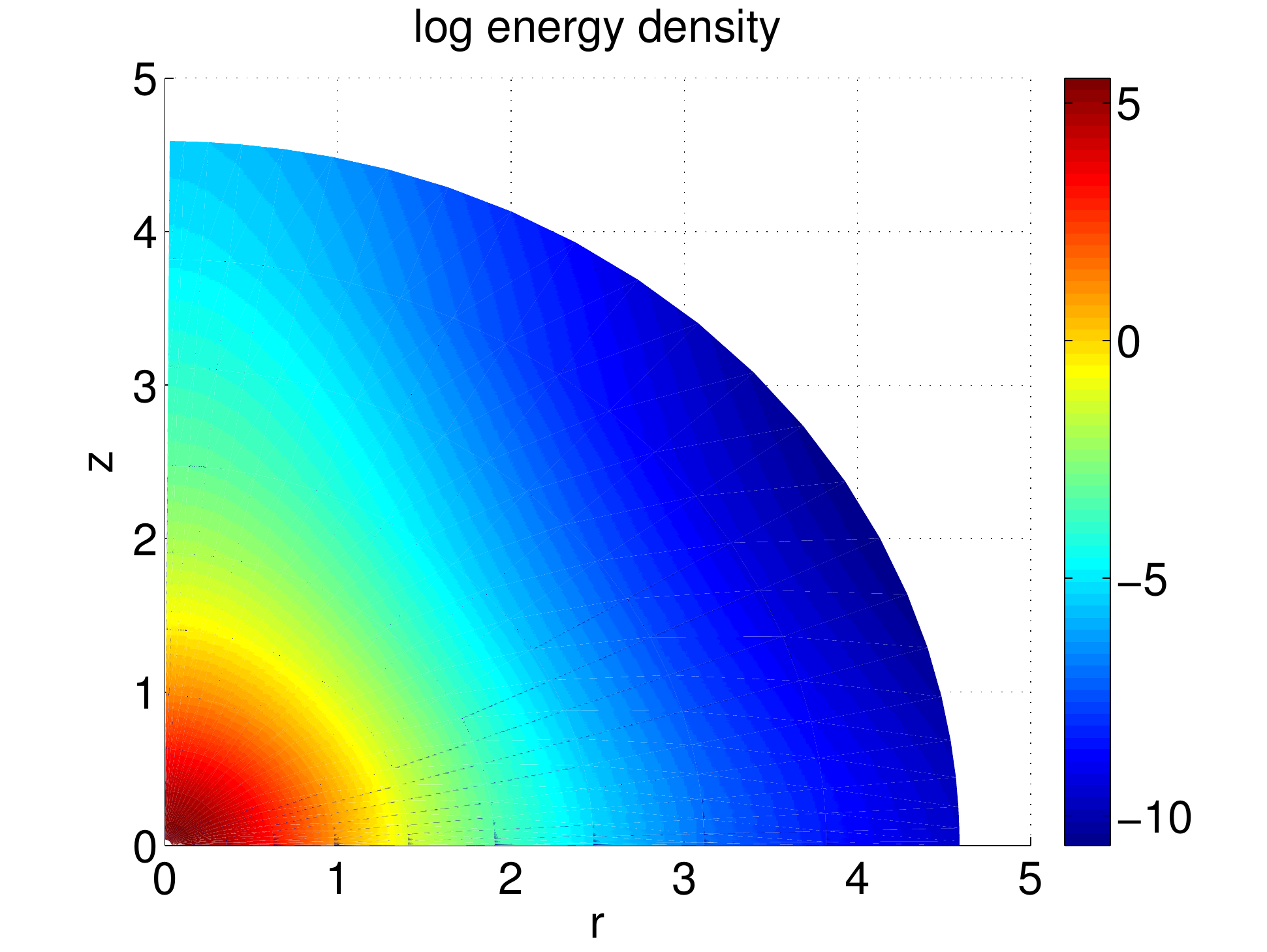}
  \caption{$\g=0.2$.} % Actual gamma is twice the saved value.
  \label{LogRhoE_02}
\end{subfigure}
\begin{subfigure}{.45\textwidth}
  \centering
  \includegraphics[width=1\linewidth]{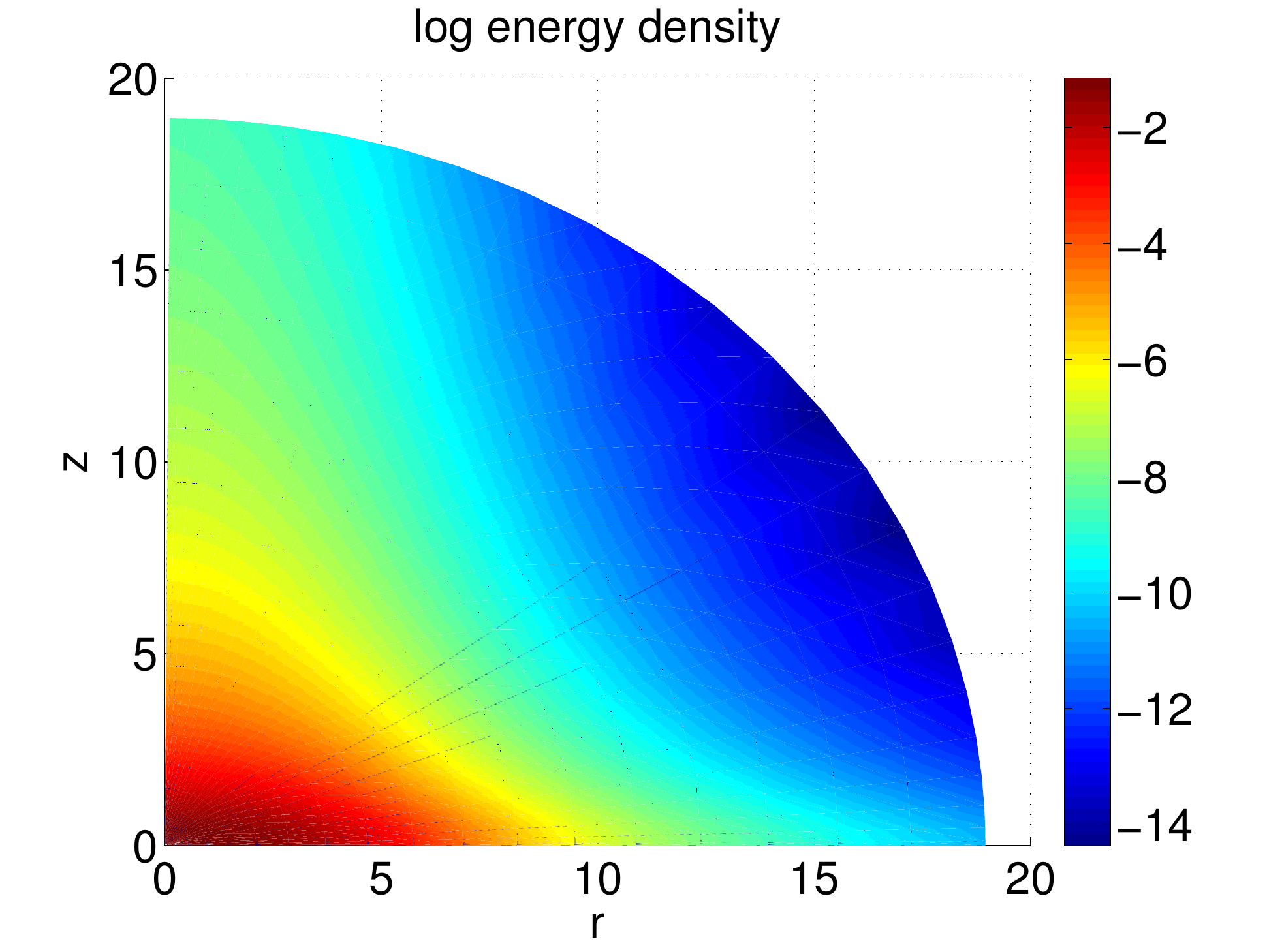}
  \caption{$\g=10$.}
  \label{LogRhoE_10}
\end{subfigure}
\caption{The logarithm of the energy density $\rho_E(r,z)$ in the $(r,z)$ plane, on the same domain as the corresponding plots in Figure \ref{RhoE_Gamma}. A large portion of the energy away from the soliton core is contained in the tail at large holographic radial coordinate $z$ and small field theory coordinate $r$.}
\label{LogRhoE_Gamma}
\end{figure}

The energy distribution of the soliton tells us how the structure is deformed as we increase the repulsion of the instanton charges by tuning the coupling $\g$. Writing the mass-energy as\footnote{We define $\rho_E(r,z)/4\pi$ as the integrand of equation (\ref{eq:Mass}) multiplied by a suitable Jacobian factor.} \be M=\frac{1}{4\pi}\int d^4x \; \rho_E(r,z), \ee we plot the energy density $\rho_E(r,z)$ of the soliton in Figures \ref{RhoE_Gamma} and \ref{LogRhoE_Gamma}. For small $\gamma$, the core of the soliton appears spherically symmetric in the $(r,z)$ plane. A closer inspection reveals a skewed tail with a slower falloff of energy density in the $z$ direction; compare Figures \ref{RhoE_02} and \ref{LogRhoE_02}. As we increase $\g$, the core of the soliton expands and deforms, smearing along the $z$-axis.

\begin{figure}[htb!]
\centering%
\includegraphics[scale=1.5]{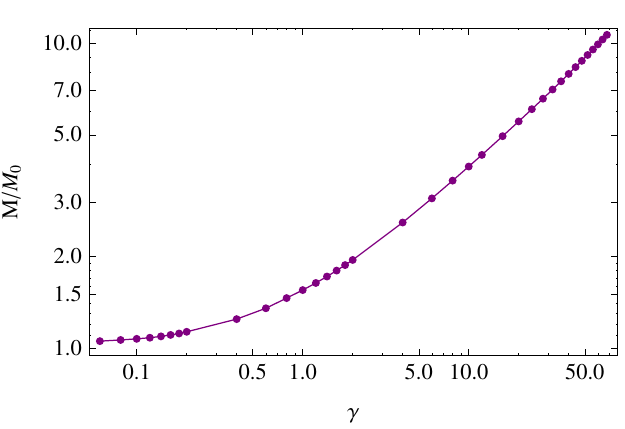}
\caption{The total mass of the soliton as a function of $\g$, normalized by the mass $M_0=8\pi^2\kappa$ of a $D4$ brane wrapping the sphere directions (equivalently the mass of a point-like $SO(4)$ instanton at $\g=0$ in the effective theory). As $\g$ decreases, the mass of the numerical solution approaches that of the point-like instanton. For $\g>10$, our results can be approximated by the relation $M\propto\g^{0.53}$.}
\label{Mass}
\end{figure}

In \cite{Sakai:2004cn}, the mass of the baryon was approximated as the energy of a D4 brane wrapping the $S^4$, giving $M_0=8\pi^2\kappa$. The mass of the wrapped D4 brane coincides with the mass of a point-like $SO(4)$ instanton at $\g=0$. By allowing a finite size spherical instanton, \cite{Hata:2007mb} computed a correction to this, finding \be M_{SO(4)} =M_0+\sqrt{\frac{2}{15}}N_c.\ee  In Figure \ref{Mass}, we plot the total mass-energy, normalized by $M_0$, of the soliton found here using the more general $SO(3)$ ansatz. As $\g$ decreases and the soliton shrinks, the effect of the curved background becomes less important and the energy approaches that of the point-like spherical instanton. As $\g$ increases and the soliton becomes more deformed, the energy of the configuration also increases. For $\g>10$, we notice that the mass-energy appears to be controlled by a power law. The best fit in this region gives $M\propto\g^{0.53}$.
 
By fitting the Sakai-Sugimoto model to the experimental values for the $\rho$ meson mass and the pion decay constant, one can fix both the parameter $\kappa$ and the energy scale in the field theory. In \cite{Hashimoto:2008zw}, this procedure yields $\kappa=0.00745$ and an energy scale such that 1 in the dimensionless units we have been using corresponds to $949$ MeV. With $N_c=3$, this gives $\gamma=2.55$. We can compare our numerical results for the baryon mass to those of the $SO(4)$ approximation for these values of the parameters. We find
\begin{align}
M_{SO(4)} &\simeq 1.60\;\textrm{GeV}, \nn\\
M_{SO(3)} &\simeq 1.19\;\textrm{GeV}.
\end{align}
There is a large difference in the results of the two approaches. Interestingly, the $SO(3)$ result is a much better approximation of the true mass of the nucleons.

\subsection{The baryon charge}
\label{sec:RhoB}

\begin{figure}
\centering
\begin{subfigure}{.45\textwidth}
  \centering
  \includegraphics[width=1\linewidth]{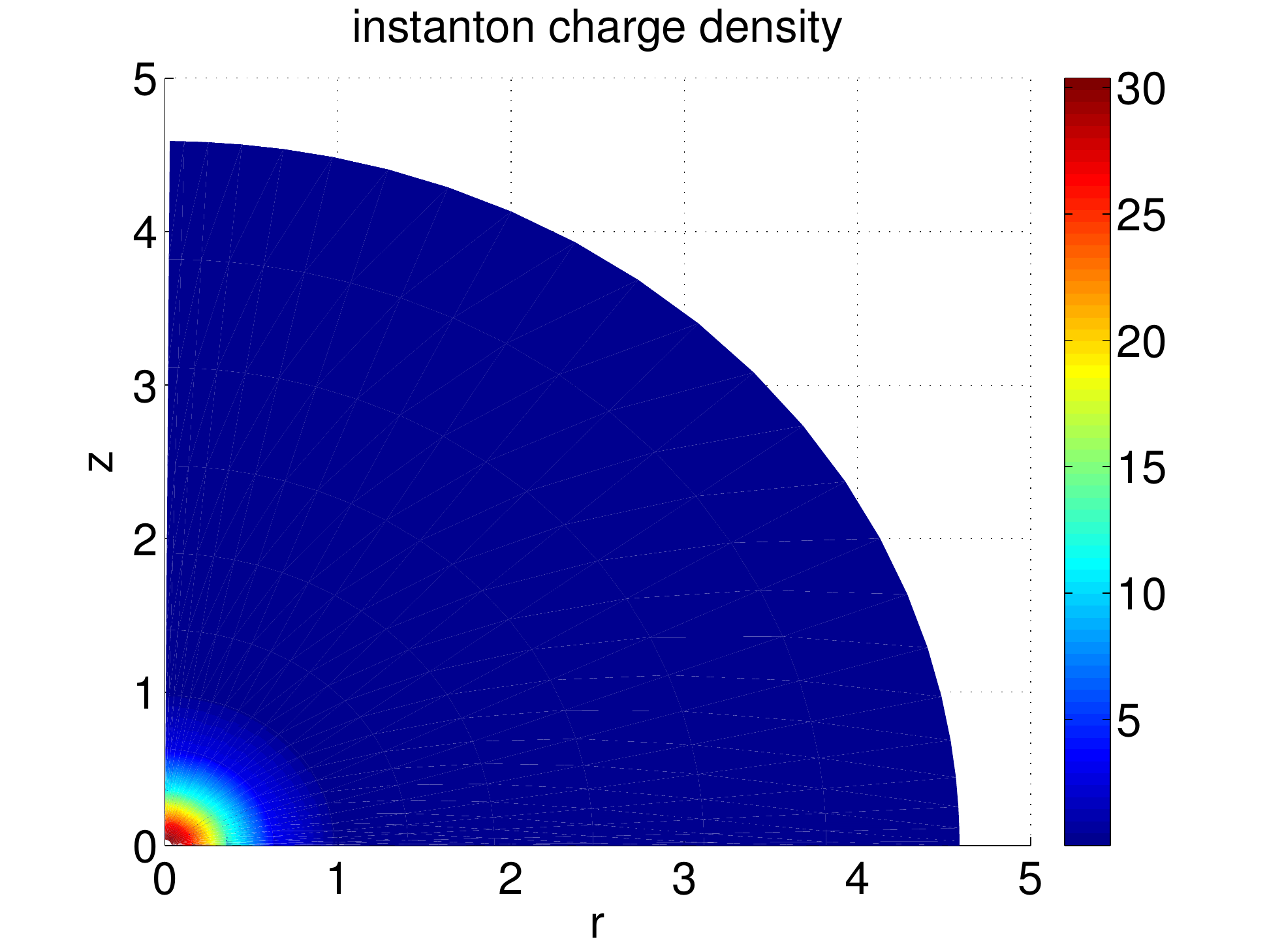}
  \caption{$\g=0.2$.} % Actual gamma is twice the saved value.
  \label{Qins_02}
\end{subfigure}
\begin{subfigure}{.45\textwidth}
  \centering
  \includegraphics[width=1\linewidth]{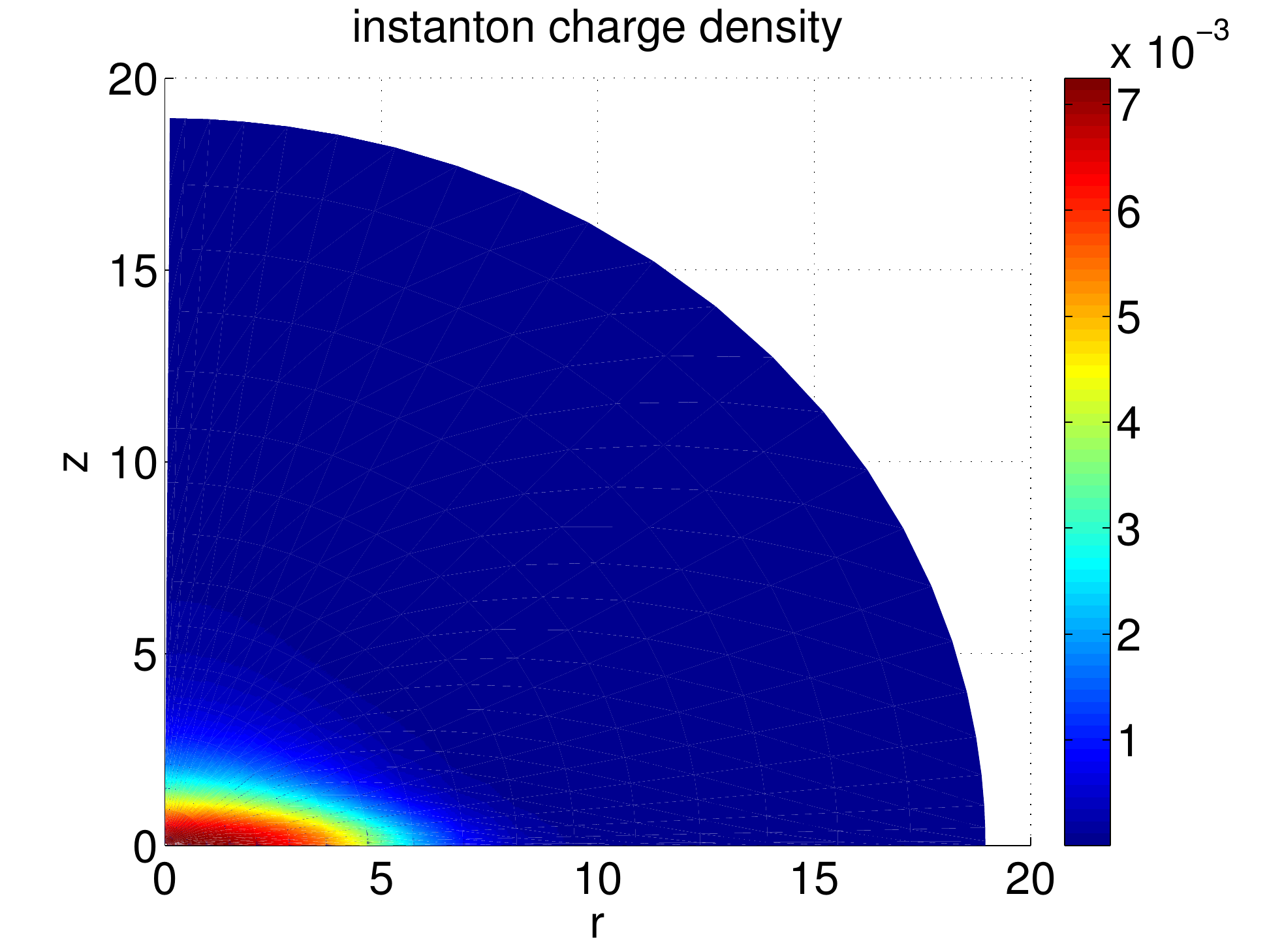}
  \caption{$\g=10$.}
  \label{Qins_10}
\end{subfigure}
\caption{The instanton number density $\frac{1}{8\pi^2} \mathrm{tr}\,F\wedge F$ in the $(r,z)$ plane. The distribution of the instanton charge closely mimics the distribution of energy density, as shown in Figures \ref{RhoE_Gamma} and \ref{LogRhoE_Gamma}.}
\label{Qins_Gamma}
\end{figure}

\begin{figure}
\centering
\begin{subfigure}{.45\textwidth}
  \centering
  \includegraphics[width=1\linewidth]{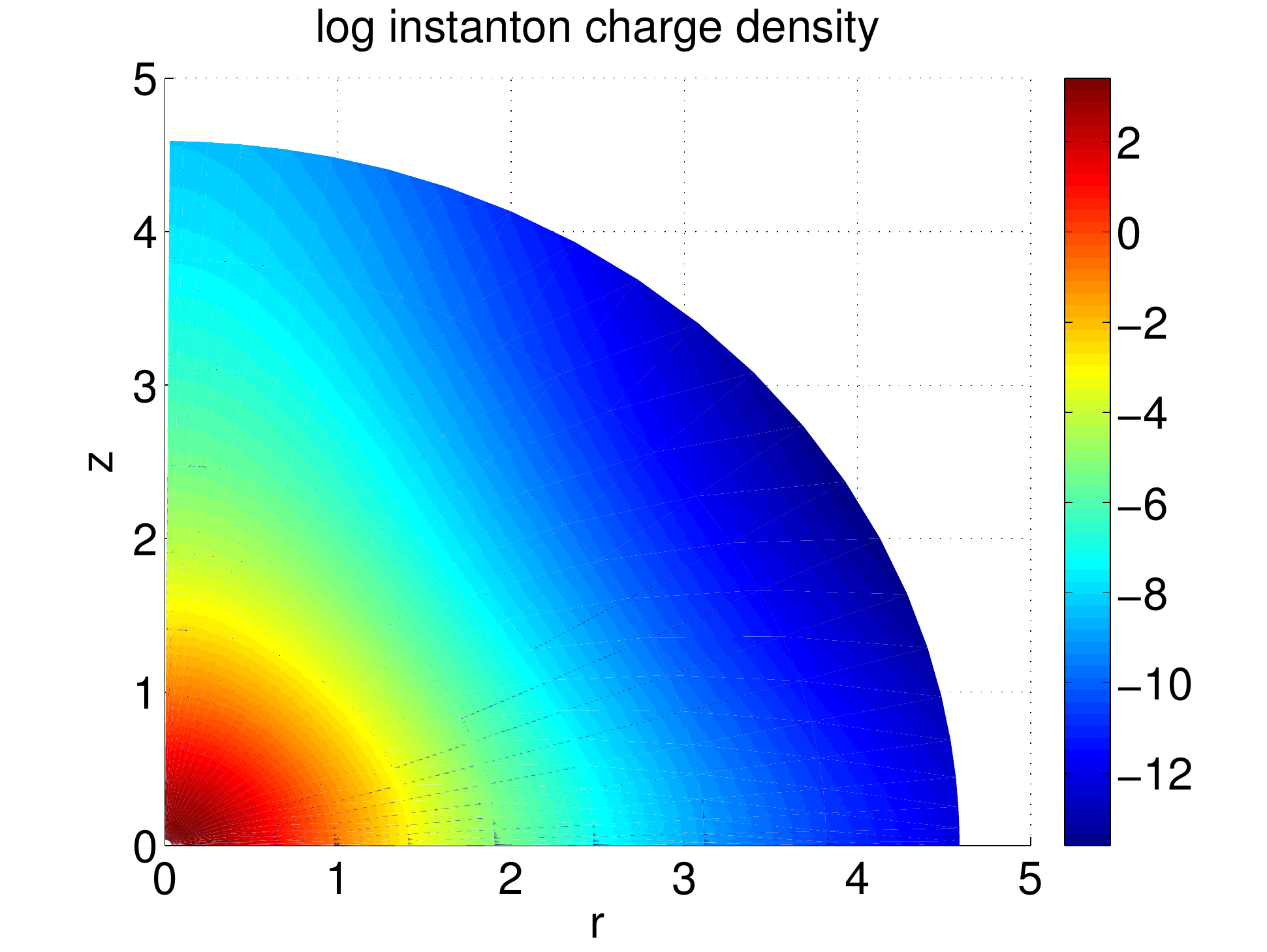}
  \caption{$\g=0.2$.} % Actual gamma is twice the saved value.
  \label{LogQins_02}
\end{subfigure}
\begin{subfigure}{.45\textwidth}
  \centering
  \includegraphics[width=1\linewidth]{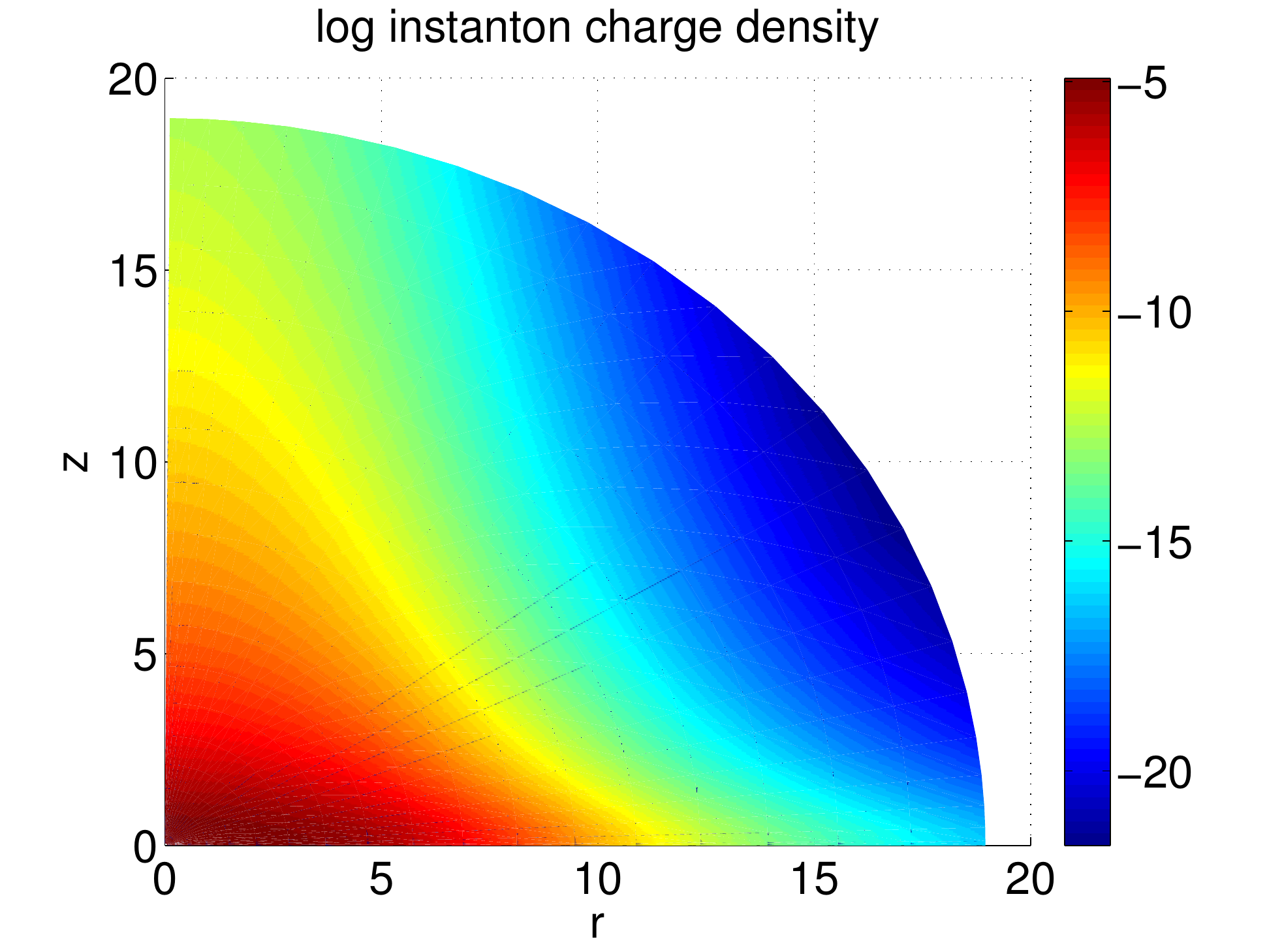}
  \caption{$\g=10$.}
  \label{LogQins_10}
\end{subfigure}
\caption{The logarithm of the instanton number density $\frac{1}{8\pi^2} \mathrm{tr}\,F\wedge F$ in the $(r,z)$ plane, on the same domain as the corresponding plots in Figure \ref{Qins_Gamma}.}
\label{LogQins_Gamma}
\end{figure}

The baryon charge in the field theory is related to the instanton number density $\frac{1}{8\pi^2} \mathrm{tr}\,F\wedge F$ in the bulk. In Figures \ref{Qins_Gamma} and \ref{LogQins_Gamma} we plot the instanton charge density for two representative solutions. The result closely matches the energy density of the soliton.

The baryon charge density can be found from the baryon number current, as defined for example in \cite{Hashimoto:2008zw}: \be J^\m_B=-\frac{2}{N_c}\kappa \(k(z)\hat{F}^{\m z}\)\Big|_{z=-\infty}^{z=\infty}.\ee Writing the Abelian gauge field near the boundary as \be\label{eq:A0} \hA_0=\frac{\hA_0^{(1)}(r)}{z}+\dots,\ee where $\dots$ denotes terms at higher order in $1/z$, we find that the baryon density is \be \rho_B(r)= J_B^0(r)= \frac{\hA_0^{(1)}(r)}{8\pi^2\g}. \ee In terms of the density, the total baryon charge is \be\label{eq:NB2} N_B=\int_0^\infty dr\, 4\pi r^2\,\rho_B(r). \ee

We fit our numerical solutions to the functional form in equation (\ref{eq:A0}) and read off the coefficient $\hA_0^{(1)}(r)$ in order to find $\rho_B(r)$. This fit is only robust up to a value of $r$ that depends on the coupling $\g$: $r=\bar{r}(\g)$. As demonstrated in \cite{Cherman:2011ve}, the charge density $\rho_B(r)$ decays as $1/r^9$. Thus the field $\hA_0$ is decaying much faster in the field theory $r$ direction than the holographic radial $z$ direction. Since we solve in the coordinate $R=(r^2+z^2)^{1/2}$, and choose a large cutoff $R_\infty$ such that the $z$ falloff is reliable, we might expect the fit to break down at some point, after $\rho_B(r)$ has decayed to a very small value. Numerically, we determine $\bar{r}(\g)$ as the point at which the error in the fit reaches ten times the error in the fit at $r=0$.

\begin{figure}[htb!]
\centering%
\includegraphics[width=\linewidth]{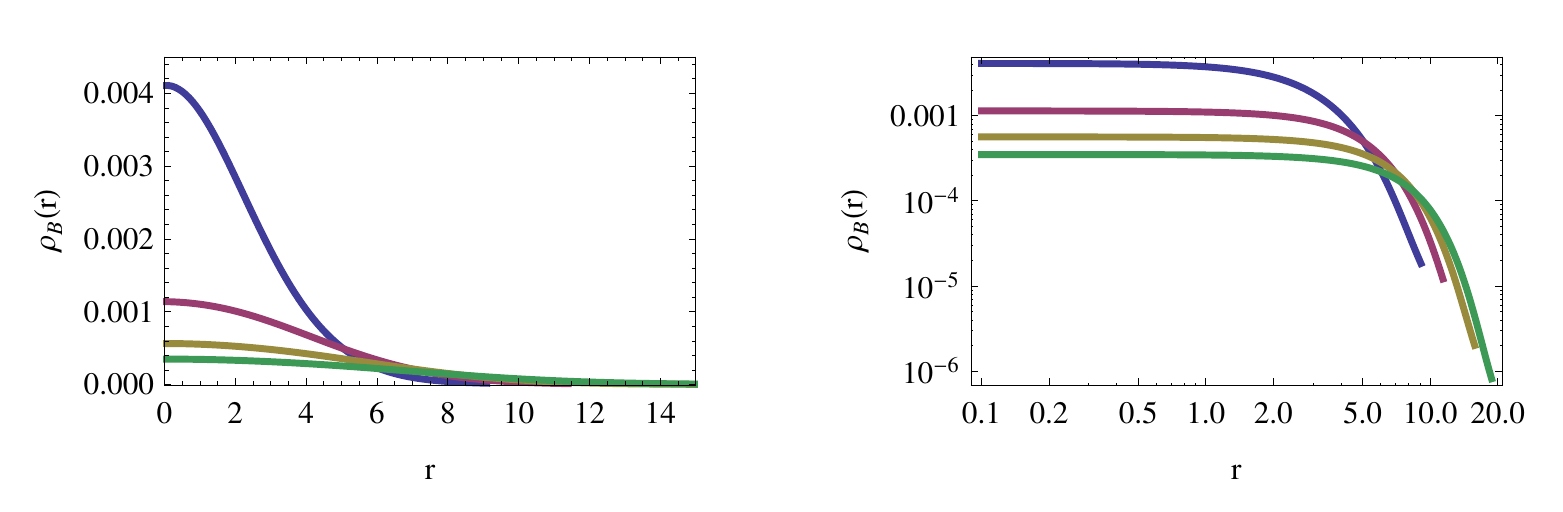}
\caption{{\it Left}: The charge density $\rho_B(r)$ for $\g=4,12,20,28$, from top to bottom. {\it Right}: The same data on a log-log axis. As $\g$ increases, the charge density becomes less peaked near the origin. The $1/r^9$ falloff of $\rho_B(r)$ behaviour can be seen in the tail of the charge distributions.}
\label{RhoB1}
\end{figure}

\begin{figure}[htb!]
\centering%
\hspace*{1in}
\includegraphics[scale=1]{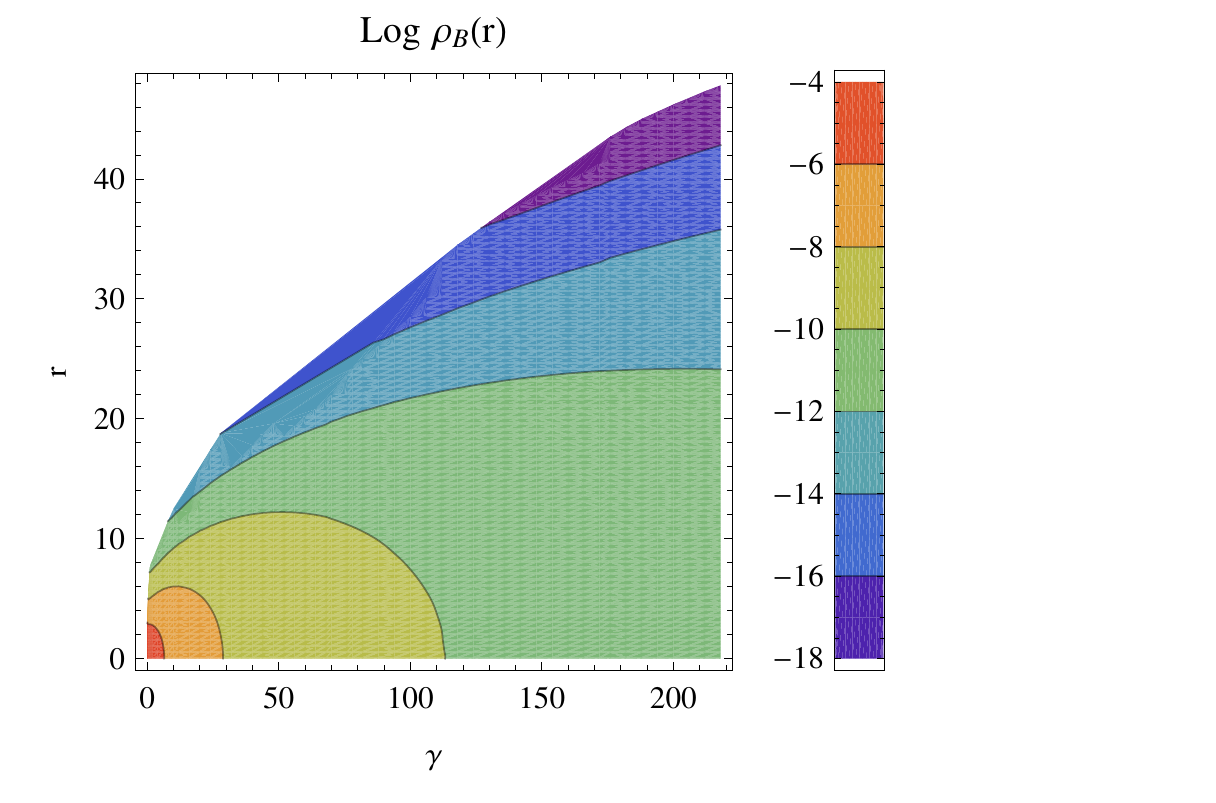}
\caption{The charge density $\rho_B(r)$ for varying $\g$.}
\label{RhoB2}
\end{figure}

In Figure \ref{RhoB1}, we plot the baryon charge $\rho_B(r)$ up to the cutoff $\bar{r}(\g)$ for various values of $\g$. As $\g$ increases, the baryon density at the origin $\rho_B(0)$ decreases and the charge moves toward the tail of the distribution. In the log-log plot, the $1/r^9$ falloff of the charge density can clearly be seen. Figure \ref{RhoB2} shows the behaviour of the baryon charge density across our entire range of $\g$.

As a check of our solution, we can compute $N_B$ by both formulas (\ref{eq:NB1}) and (\ref{eq:NB2}). We find that, across the range of $\g$ and using both formulas, $N_B=1$ to good precision.

\begin{figure}[htb!]
\centering%
\includegraphics[scale=1.5]{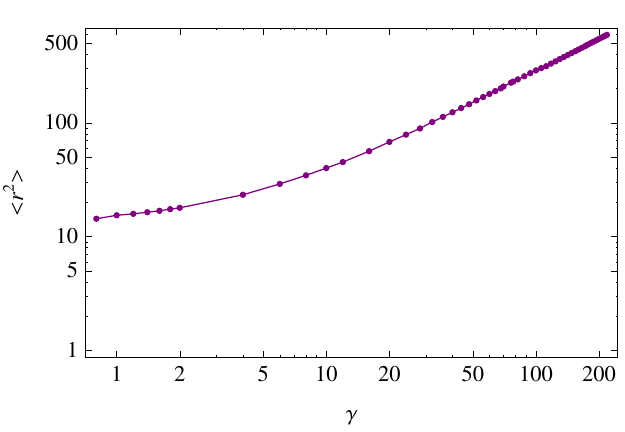}
\caption{The baryon charge radius $\langle r^2\rangle=\int r^2 \(4\pi r^2\rho_B(r)\)dr$ as a function of $\g$. For $\g>35$, the relation can be approximated by $\langle r^2\rangle\propto \g^{0.93}$.}
\label{IsoRad}
\end{figure}

Lastly, with the charge density $\rho_B(r)$, we can compute the baryon charge radius \be\langle r^2\rangle=\int_0^\infty r^2 \(4\pi r^2\rho_B(r)\)dr.\ee To integrate past the cutoff $\bar{r}(\g)$, we approximate the tail of the distribution as $\rho_B(r;\g)\sim c(\g)/r^9$, where $c(\g)$ is approximated from the value of the density at the integration cutoff. The baryon charge radius is plotted in Figure \ref{IsoRad}. For $\g>35$, the relation appears to obey a power law, with best fit given by $\langle r^2\rangle\propto \g^{0.93}$.

As above, it is interesting to compare the result to that obtained from the $SO(4)$ approximation, evaluated at the parameters defined by the fit to meson physics. The result is\footnote{We compare to the result from the classical analysis of the $SO(4)$ baryon, given in equation (3.11) of \cite{Hashimoto:2008zw}.}
\begin{align}
\langle r^2\rangle^{1/2}_{SO(4)} &\simeq 0.785\;\textrm{fm}, \nn\\
\langle r^2\rangle^{1/2}_{SO(3)} &\simeq 0.90\;\textrm{fm}.
\end{align}
In this model, the baryon charge radius equals the electric charge radius of the proton \cite{Hashimoto:2008zw}. The result from our numerics is very close to the experimental value for the electric charge radius of the proton, which has been measured to be in the range 0.84 fm -- 0.88 fm.

%====================================================================%
\section{Conclusion}
\label{sec:Conc}

We have studied properties of baryons in a holographic model of QCD related to the Sakai-Sugimoto model by simplifying the Born-Infeld part of the D8-brane action to a 5D Yang-Mills plus Chen-Simons action for the gauge fields in the non-compact directions. By dropping the assumption of $SO(4)$ symmetry and finding direct solutions to the bulk field equations for the gauge field, we have found that various properties of the baryons in the holographic QCD model change significantly. In particular, the baryon mass gives substantially better agreement with measured values. There are several interesting directions for future work.

Within the present model, it would be interesting to calculate other observables such as the form-factors associated with the isospin currents (associated with the $SU(2)$ flavour symmetry) and compare these to results calculated using the $SO(4)$ symmetric ansatz \cite{Hashimoto:2008zw}. It would also be interesting to consider interactions between two baryons. This requires a less-symmetric ansatz, but the numerics should still be feasible. Again, it would be interesting to compare with previous results calculated assuming flat-space instanton configurations \cite{Hashimoto:2009ys}. For higher baryon charge, it should be feasible to consider the question of nuclear masses as a function of baryon number, at least within the space of $SO(3)$-symmetric configurations. The actual ground states for higher baryon number may not be so symmetric however. Finally, it would be interesting to investigate solutions with a finite baryon charge density (e.g. at finite baryon chemical potential). Such configurations were considered with various simplifying assumptions in \cite{Horigome:2006xu, Sin:2007ze, Yamada:2007ys, Bergman:2007wp, Kaplunovsky:2012gb, Rozali:2007rx}. As shown in \cite{Rozali:2007rx}, these are necessarily inhomogeneous in the field theory directions, so a numerical approach similar to the one used in this paper is likely necessary to investigate detailed properties of the ground state at various densities.

Finally, it is interesting to investigate effects of replacing the Yang-Mills action used here with the full D8-brane Born-Infeld action. This is incompletely known, but one could work for example with the Abelian Born-Infeld action promoted to a non-Abelian action via the symmetrized trace prescription that has been shown to be correct for the $F^4$ terms. While the equations in this case will be significantly more complicated, they should pose no serious obstacle for the numerical approach that we are using. An interesting difference between the Born-Infeld and Maxwell actions for Abelian gauge fields is that the Maxwell action associates an infinite energy to point charges, while this energy is finite in the Born-Infeld case. Thus, we might expect that the tendency for the instantons to spread out is somewhat less with the Born-Infeld action. In this case, we may expect a somewhat smaller, less massive baryon. Thus, the baryon mass in the model using the Born-Infeld action may be even closer to the experimental value than we have found here.

\section*{Acknowledgements}

We thank Kevin Whyte for early collaboration on this project and Shigeki Sugimoto for useful discussions. M.R thanks Centro de Ciencias de Benasque Pedro Pascual, IPMU at the University of  Tokyo and INI at Cambridge University for hospitality during the course of this work. This work is supported in part by the Natural Sciences and Engineering Research Council of Canada.

%====================================================================%
\bibliography{baryon_refs}
\bibliographystyle{naturemag}

\end{document}